%% file: arxiv.tex
\documentclass[conference]{IEEEtran}
\IEEEoverridecommandlockouts
\usepackage{cite}
\usepackage{amsmath,amssymb,amsfonts}
\usepackage{algorithmic}
\usepackage{graphicx}
\usepackage{textcomp}
\usepackage{xcolor}
\usepackage{booktabs}
\usepackage{multirow}
\usepackage{url}
\def\BibTeX{{\rm B\kern-.05em{\sc i\kern-.025em b}\kern-.08em
    T\kern-.1667em\lower.7ex\hbox{E}\kern-.125emX}}

\newif\ifblind
\begin{document}

\title{Encoder Awakening via Adapters: Effective Domain-Adaptive Fine-tuning of Speech-LLMs\\
% {\footnotesize \textsuperscript{*}Note: Sub-titles are not captured for https://ieeexplore.ieee.org  and
}

\ifblind
\author{\parbox[c][6em][c]{\linewidth}{\centering\LARGE \textit{Anonymous Submission to SLT 2026}}}
\else
\author{
\IEEEauthorblockN{
Mohan Shi, 
Zilai Wang,
Natarajan Balaji Shankar,
Kaiyuan Zhang,
Eray Eren,
Abeer Alwan}

\IEEEauthorblockA{
\textit{Department of Electrical and Computer Engineering} \\
\textit{University of California Los Angeles}\\
Los Angeles, USA \\
\textit{\{shimohan, zilaiwang2001, balaji1312, kaiyuanzhang, erayeren\}@ucla.edu, alwan@ee.ucla.edu}
}
}
\fi

\maketitle

\begin{abstract}
% Speech Large Language Models (Speech-LLMs), which integrate a pre-trained speech encoder and a large language model (LLM) via a modality projector, and are fine-tuned with Low-Rank Adapters (LoRA) applied to the LLM, have demonstrated strong performance not only in automatic speech recognition (ASR) but also in a range of other tasks. However, adapting a well-trained Speech-LLM to a new target domain (e.g., child speech or dialectal speech) remains challenging due to significant domain shift and limited training data. In particular, since the LLM is the dominant component of the model, the speech encoder may not be sufficiently adapted to the new acoustic domain during fine-tuning.

% In this paper, we propose Encoder Awakening via Adapters (EAVA), a simple yet effective method for domain-adaptive fine-tuning of Speech-LLMs for ASR. EAVA comprises two stages: (1) lightweight adapters are inserted into each encoder layer and trained exclusively, allowing target-domain knowledge to be injected into the encoder while preserving its pre-trained knowledge; (2) the full model is jointly fine-tuned on the target domain, with LoRA applied to the LLM. Experiments on three domain-shifted ASR datasets, including two child speech datasets and one dialectal speech dataset, demonstrate that the proposed method significantly outperforms all baselines, including vanilla fine-tuning, and achieves new state-of-the-art performance.
Speech Large Language Models (Speech-LLMs), typically built from a pre-trained speech encoder, a modality projector, and an LLM fine-tuned with Low-Rank Adapters (LoRA), have shown strong Automatic Speech Recognition (ASR) performance on general-domain speech. However, adapting them to domain-shifted speech, such as child or dialectal speech, remains challenging under limited target-domain data. Given the dominant role of the LLM in Speech-LLMs, with cross-entropy loss applied only at the LLM output, the speech encoder may receive insufficient adaptation to new acoustic conditions.

In this paper, we propose Encoder Awakening via Adapters (EAVA), a simple yet effective domain-adaptive fine-tuning method for Speech-LLM-based ASR. First, lightweight adapters are inserted into each encoder layer and trained exclusively, enabling target-domain acoustic knowledge to be incorporated into the encoder while preserving its pre-trained knowledge. Second, the full model is jointly fine-tuned on the target domain with LoRA applied to the LLM. Experiments on three domain-shifted ASR datasets, covering child and dialectal speech, show that EAVA consistently outperforms vanilla fine-tuning and other baselines, achieving new state-of-the-art performance.\footnote{Code is available at \url{https://github.com/morganshi/EAVA}}
\end{abstract}


\input{introduction}

\input{method}

\input{experiments}

\input{conclusions}

\section{Acknowledgements}
\label{sec:ack}
This research is supported in part by the National Science Foundation (NSF) and the Institute of Education Sciences (IES), U.S. Department of Education (DoE), through Grant R305C240046 to the U. at Buffalo. The opinions expressed are those of the authors and do not represent views of the IES, DoE, or the NSF.

\newpage

\section{Generative AI Use Disclosure}
\label{sec:genai_disclosure}
During the preparation of this work, the authors used ChatGPT (GPT-5.5) for language editing, including proofreading and improving clarity and readability of the manuscript. All technical content, experimental design, results, and conclusions were produced and verified by the authors. After the use of Generative AI, the authors reviewed and edited the manuscript and take full responsibility for the content of the publication. Generative AI tools were not used to produce a significant portion of the manuscript and are not listed as authors.


% \vspace{12pt}
% \color{red}
% IEEE conference templates contain guidance text for composing and formatting conference papers. Please ensure that all template text is removed from your conference paper prior to submission to the conference. Failure to remove the template text from your paper may result in your paper not being published.

\end{document}

%% file: introduction.tex
\section{Introduction}
Speech Large Language Models (Speech-LLMs), which benefit from a pre-trained speech encoder~\cite{BaevskiZMA20,HsuBTLSM21,ChenWCWLCLKYXWZ22} or tokenizer~\cite{defossez2022high,zhang2023speechtokenizer} and the language generation and reasoning capabilities of powerful Large Language Models (LLMs)~\cite{grattafiori2024llama,abdin2024phi,yang2025qwen3}, have shown strong performance on a wide range of speech-related tasks~\cite{FathullahWLJSLG24,ma2024embarrassingly,ShiJXXZWSZY24,chen2025neural,TangYSC000M024,wang2026emotionthinker,shi2026train}. A standard Speech-LLM for Automatic Speech Recognition (ASR) leverages a pre-trained speech encoder, a modality projector, and a powerful LLM, fine-tuned together on large-scale paired speech-transcription data with Low-Rank Adaptation (LoRA)~\cite{HuSWALWWC22} applied to the LLM. This framework has shown impressive ASR performance on general-domain speech~\cite{nvidia_canary_qwen_25b,microsoft_phi4_multimodal}.

However, for speech domains with substantial domain shift relative to the general domain, such as child speech~\cite{ShobakiHC00,PradhanCW24} and dialectal speech~\cite{kendall2023coraal}, pre-trained Speech-LLMs often perform poorly due to the acoustic mismatch. Simply fine-tuning the encoder, projector and LoRA in LLM using cross-entropy (CE) loss is the standard way to adapt a well-trained Speech-LLM to a target domain for ASR task, but it is not effective enough for handling the acoustic domain shift. For example, no work has shown Speech-LLMs can improve performance on children ASR; the current state-of-the-art results are still from fine-tuning other speech foundation models (not Speech-LLMs)~\cite{FanSA24,ying25_wocci,wang2026gumbelbeard}. One challenge in domain-adaptive fine-tuning of Speech-LLMs is that the speech encoder may remain insufficiently adapted to the target acoustic domain, as the large LLM component tends to dominate the model and the cross-entropy loss is applied only at the LLM output. Moreover, such domain-shift scenarios typically involve limited data for fine-tuning, further complicating adaptation~\cite{ShankarFA24}. On the other hand, if the encoder is trained too aggressively, it may fail to preserve its pre-trained knowledge, leading to degraded performance. So how to effectively adapt a pre-trained Speech-LLM to a target acoustic domain is a challenging and underexplored problem. Other works adopt carefully designed multi-stage alignment training procedures to better align the speech and LLM representation spaces.~\cite{ShiJXXZWSZY24,MuSWYX25}. Although these approaches achieve competitive performance in their settings, their objective is to integrate a raw speech encoder with an LLM from scratch to build a Speech-LLM, which is fundamentally different from adapting a well-trained model to a new acoustic domain. Moreover, their procedures are complex and may not generalize to domain-adaptive fine-tuning task.

To address this gap, we propose Encoder Awakening via Adapters (EAVA), a simple yet effective method for domain-adaptive fine-tuning of Speech-LLMs for ASR. EAVA consists of two stages. In the first stage, termed Encoder Awakening, lightweight adapters are inserted into each layer of the pre-trained speech encoder, and only these adapters are trained while all other components remain frozen. This design injects target-domain acoustic knowledge into the encoder while preserving its pre-trained capabilities. In the second stage, termed Continual Fine-tuning, the encoder adapters, the projector, and the LoRA modules in the LLM are jointly fine-tuned to further adapt the Speech-LLM to the target domain, leveraging the domain-aware initialization obtained from the first stage. Experiments on several representative domain-shifted datasets, including child and dialectal speech, show that EAVA significantly outperforms all baselines and achieves new state-of-the-art ASR results on all evaluated datasets. The method also demonstrates consistent improvements across different Speech-LLM backbones.

%% file: method.tex
\section{Background and Related Works}
\subsection{Speech Large Language Models for ASR}
\begin{figure}[t]
    \centering
    \includegraphics[width=0.88\columnwidth]{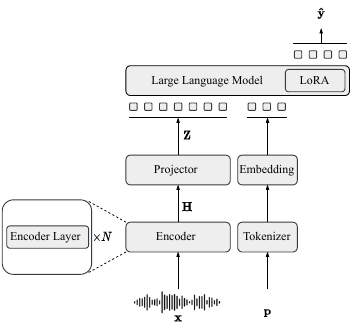}
    \vspace{-0.2cm}
    \caption{Illustration of a standard well-trained Speech-LLM for ASR. LLM autoregressively generates the transcription conditioned on embedded speech representation and the embedded prompt.}
    \vspace{-0.1cm}
    \label{fig:speech_llm}
\end{figure}

Speech-LLMs combine a pre-trained speech encoder with a powerful LLM via a modality projector to enable end-to-end Automatic Speech Recognition (ASR)~\cite{FathullahWLJSLG24}, as illustrated in Figure~\ref{fig:speech_llm}. The speech encoder, typically based on the Conformer~\cite{GulatiQCPZYHWZW20} or similar architectures, extracts frame-level acoustic representations from the input speech signal. The projector maps these representations into the token embedding space of the LLM, which then generates text output autoregressively. These components are jointly fine-tuned, with Low-Rank Adaptation (LoRA) commonly applied to the LLM. Formally, given a speech signal $\mathbf{x}$ and a text prompt $\mathbf{p}$, the forward process is:
\begin{align}
    \mathbf{H} &= \text{Encoder}(\mathbf{x}) \in \mathbb{R}^{T \times d_e}, \label{eq:enc} \\
    \mathbf{Z} &= \text{Projector}(\mathbf{H}) \in \mathbb{R}^{T \times d_l}, \label{eq:proj}
\end{align}
where $T$ is the sequence length, and $d_e$, $d_l$ are the encoder and LLM hidden sizes, respectively. The LLM then autoregressively generates the transcription $\mathbf{\hat{y}} = (\hat{y}_1, \ldots, \hat{y}_N)$ conditioned on $\mathbf{Z}$ and the embedded prompt: $\text{Embedding}(\text{Tokenizer}(\mathbf{p}))$.

Representative open-source Speech-LLMs include Canary-Qwen~\cite{nvidia_canary_qwen_25b}, which pairs a pre-trained FastConformer encoder~\cite{RekeshKKMNHHPKBG23} with a decoder-only LLM Qwen3-1.7B~\cite{yang2025qwen3}, and Phi-4-Multimodal~\cite{microsoft_phi4_multimodal}, which integrates speech understanding into the Phi-4-Mini~\cite{abouelenin2025phi} language model via a dedicated speech encoder. These models are pre-trained on large-scale paired speech-transcription data and achieve strong ASR performance on general-domain speech.

\subsection{Problem Formulation}
Consider a Speech-LLM pre-trained on a large general-domain dataset, which generates ASR transcriptions from speech inputs. Given a target-domain training set $\mathcal{D}_\text{tgt}^\text{train}$ exhibiting acoustic domain shift (e.g., child or dialectal speech), domain-adaptive fine-tuning aims to adapt the model for improved ASR performance on the target domain.

\subsection{Previous Fine-tuning Strategies}
\subsubsection{Vanilla Fine-tuning}
\label{standard_ft}
The most straightforward approach is vanilla fine-tuning, which involves jointly fine-tuning the encoder, projector, and LLM LoRA on the target-domain training set $\mathcal{D}_\text{tgt}^\text{train}$ using the cross-entropy (CE) loss computed between the predicted and ground-truth transcriptions. However, as discussed in the Introduction, this approach struggles to balance effective encoder adaptation and preservation of pre-trained knowledge under limited target-domain data.

\subsubsection{Multi-stage Alignment}
\label{multi_ft}
Previous works~\cite{ShiJXXZWSZY24,MuSWYX25} have adopted multi-stage alignment for Speech-LLMs to better align the speech and LLM representation spaces. Specifically, \cite{MuSWYX25} proposed a training strategy that first fine-tuned the encoder, then trained only the projector, and finally trained the projector and LLM LoRA jointly. However, their objective is to integrate a raw speech encoder with an LLM from scratch, rather than to adapt a well-trained model to a new domain.

\begin{figure*}[t]
    \centering
    \includegraphics[width=\textwidth]{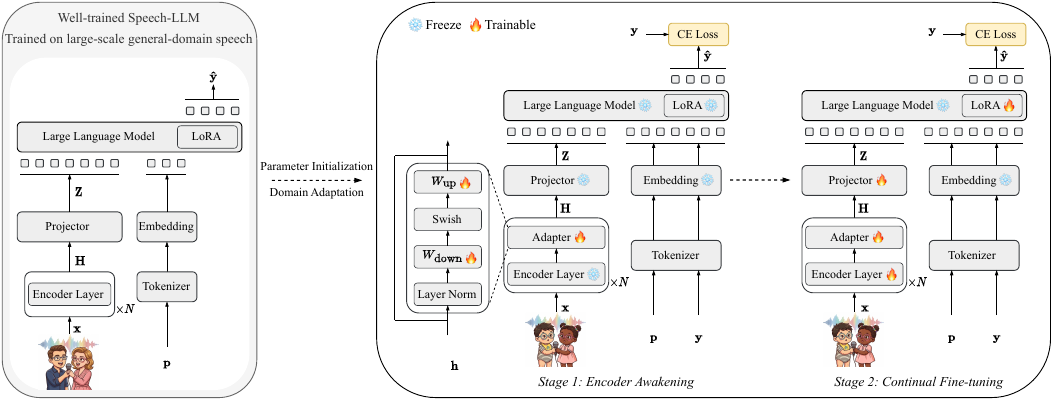}
    \caption{Overview of the proposed EAVA method for domain-adaptive fine-tuning of a well-trained Speech-LLM. A Speech-LLM trained on large-scale general-domain speech is used as the parameter initialization and adapted to target-domain speech, such as child or dialectal speech. In Stage 1 (Encoder Awakening), lightweight adapters are inserted into each encoder layer and only the adapter parameters are trained, while all other components remain frozen. In Stage 2 (Continual Fine-tuning), the encoder adapters, the pre-trained encoder parameters, the projector, and the LLM LoRA modules are jointly fine-tuned.}
    \vspace{-0.2cm}
    \label{fig:eava}
    \vspace{-0.1cm}
\end{figure*}

\section{Method}
As discussed in the previous sections, effective domain-adaptive fine-tuning strategies for Speech-LLMs remain largely unexplored. To address this gap, we propose Encoder Awakening via Adapters (EAVA), a simple yet effective method for this task. Figure~\ref{fig:eava} illustrates an overview of EAVA, which consists of two stages.

\subsection{Stage 1: Encoder Awakening}
The first stage is the encoder awakening stage, which aims to awaken the encoder toward the target domain. Specifically, starting from a well-trained Speech-LLM, randomly initialized lightweight adapters are inserted into each encoder layer, and only these adapters are trained on the target-domain training set $\mathcal{D}_\text{tgt}^\text{train}$ using the standard LLM CE loss, while all pre-trained parameters remain frozen, as shown in Figure~\ref{fig:eava} (Stage 1). This stage aims to effectively inject target-domain knowledge into the encoder via the adapters, while preserving the pre-trained knowledge of the encoder itself. In this work, we employ Residual Adapters (RA)~\cite{BapnaF19} as the default adapter choice, each consisting of two feed-forward layers with a residual connection, along with layer normalization and a Swish activation function~\cite{ramachandran2017swish}. Formally, let $\mathbf{h} \in \mathbb{R}^{d_e}$ be the output hidden state of an encoder layer. The RA computes:
\begin{equation}
    \text{RA}(\mathbf{h}) = \mathbf{h} + W_\text{up} \cdot \text{Swish}\!\left(W_\text{down} \cdot \text{LayerNorm}(\mathbf{h})\right)
    \label{eq:ra}
\end{equation}
where $W_\text{down} \in \mathbb{R}^{r \times d_e}$ and $W_\text{up} \in \mathbb{R}^{d_e \times r}$ are the adapter weights, and $r$ is the bottleneck dimension (referred to as the adapter dimension in experiments). During Stage 1, only $\{W_\text{down}, W_\text{up}\}$ in each encoder layer are updated; all other parameters remain frozen.

\subsection{Stage 2: Continual Fine-tuning}
% After the Encoder Awakening stage, the encoder adapters, the pre-trained encoder parameters, the projector, and the LLM LoRA modules are jointly fine-tuned on $\mathcal{D}_\text{tgt}^\text{train}$ using the same CE loss, as shown in Figure~\ref{fig:eava} (Stage 2). The encoder adapters, having been initialized with target-domain acoustic knowledge in Stage 1, provide a domain-aware starting point for the encoder. Joint fine-tuning then allows the projector and the LLM to align with the updated encoder representations, resulting in coherent adaptation across the entire model. This two-stage design decouples the acoustic adaptation of the encoder from the broader cross-modal alignment update, making the optimization in each stage more targeted compared to directly performing joint fine-tuning.
After the Encoder Awakening stage, the encoder adapters, the pre-trained encoder parameters, the projector, and the LLM LoRA modules are jointly fine-tuned on $\mathcal{D}_\text{tgt}^\text{train}$ using the same CE loss, as shown in Figure~\ref{fig:eava} (Stage 2). The encoder adapters, having been initialized with target-domain acoustic knowledge in Stage 1, provide a domain-aware starting point for the encoder. Joint fine-tuning then further adapts the entire Speech-LLM to the target domain, allowing the encoder, projector, and LLM LoRA modules to be updated coherently under the same ASR objective. This two-stage design first effectively awakens the encoder toward the target acoustic domain while preserving its pre-trained knowledge, and then refines the entire Speech-LLM for target-domain ASR, resulting in more targeted adaptation than directly fine-tuning all trainable components jointly.

%% file: experiments.tex
\begin{table*}[t]
\centering
\caption{WER (\%) on three target-domain datasets using Canary-Qwen as the backbone. Adapter parameters are reported as percentages of the encoder size (810M). Bold denotes the best result in each column. All EAVA variants achieve statistically significant improvements over all three baselines implemented in our experiments (Zero-shot, Vanilla Fine-tuning, and Multi-stage Alignment), as {measured by MAPSSWE} ($p < 0.05$).}
\vspace{-0.2cm}
\label{tab:main}
\begin{tabular}{lcc cc cc cc}
\toprule
\multirow{2}{*}{Method} & \multirow{2}{*}{Adapter Dim} & \multirow{2}{*}{Adapter Params} & \multicolumn{2}{c}{OGI} & \multicolumn{2}{c}{MyST} & \multicolumn{2}{c}{CORAAL} \\
\cmidrule(lr){4-5} \cmidrule(lr){6-7} \cmidrule(lr){8-9}
& & & dev & test & dev & test & dev & test \\
\midrule
Previous SOTA~\cite{FanSA24,ying25_wocci,shankar2026gclora}          & -- & -- & 10.40 & 11.60 & 7.90 & 8.50 & --   & 9.70  \\
\quad + Additional Unlabeled Training Data~\cite{wang2026gumbelbeard}            & -- & -- & --    & 11.06 & 7.74 & 8.21 & 6.10 & 9.25  \\
Zero-shot Canary-Qwen & --   & --    & 14.28 & 16.30 & 8.24 & 8.96 & 8.93 & 12.37 \\
Vanilla Fine-tuning            & --   & --    & 9.39  & 10.95 & 7.63 & 8.37 & 6.11 & 8.87  \\
Multi-stage Alignment~\cite{MuSWYX25} & --   & --    & 9.29  & 10.79 & 7.86 & 8.60 & 6.50 & 9.54  \\
\midrule
\multirow{5}{*}{EAVA (Ours)} & 32   & 2.2M (0.27\%)  & \textbf{7.85} & 9.32 & 7.35 & 8.07 & \textbf{5.84} & \textbf{8.53} \\
                                  & {64}   & {4.3M (0.53\%)} & {8.15} & \textbf{{9.05}} & \textbf{{7.33}} & \textbf{{8.04}} & \textbf{{5.84}} & {8.54} \\
                                  & 128  & 8.5M (1.05\%)  & 8.13 & 9.59 & 7.35 & \textbf{8.04} & 5.86 & 8.55 \\
                                  & 256  & 16.8M (2.07\%) & 8.19 & 9.34 & 7.44 & 8.09 & 5.87 & 8.67 \\
                                  & 512  & 33.6M (4.15\%) & 8.50 & 9.81 & 7.40 & 8.11 & 5.89 & 8.67 \\
                                %   & 1024 & --    & 8.16 & 9.34 & --   & --   & 5.86 & 8.74 \\
\bottomrule
\end{tabular}
\vspace{-0.2cm}
\end{table*}

\section{Experimental Settings}
\subsection{Datasets}
To evaluate our proposed method, we conduct experiments on three datasets that exhibit acoustic domain shift from the general adult English speech domain, covering child speech and dialectal speech variations in english.

\begin{itemize}
    \item \textbf{OGI}~\cite{ShobakiHC00}: The OGI Kids corpus contains child speech collected in classroom settings. We select its \textbf{spontaneous} portion, which consists of responses to open-ended questions from children aged 4--15, with transcriptions that preserve disfluencies. We refer to this subset as OGI throughout the paper. Following~\cite{ying25_wocci}, we split OGI into 22/2/7 hours for train/dev/test.

    \item \textbf{MyST}~\cite{PradhanCW24}: The MyST corpus comprises dialogues between elementary school students aged 8--10 and virtual tutors. We split the transcribed portion following~\cite{FanSA24,AttiaL0DE24}, resulting in 133/21/25 hours for train/dev/test.

    \item \textbf{CORAAL}~\cite{kendall2023coraal}: The CORAAL corpus contains sociolinguistic interviews in African American Language. Following ~\cite{wang2026gumbelbeard,shankar2026gclora}, we use six subsets (ATL, LES, DCA, DCB, DTA, PRV; 137h) for training, and hold out ROC (13h) and VLD (12h) for development and testing, ensuring speaker and regional disjointness. Utterances are trimmed to retain only interviewee speech and are capped at 30 seconds.
\end{itemize}

\subsection{Model Settings}
For the main experiments, we adopt Canary-Qwen~\cite{nvidia_canary_qwen_25b} as the primary backbone. Canary-Qwen is an open-source Speech-LLM developed by NVIDIA that pairs a pre-trained FastConformer encoder~\cite{RekeshKKMNHHPKBG23} (810M parameters) with a Qwen3-1.7B LLM~\cite{yang2025qwen3}. It is pre-trained on large-scale multilingual speech-transcription data and achieves strong performance on general-domain ASR benchmarks. For the adapter dimension, we use 64 by default and also evaluate other sizes: 32, 128, 256, and 512. For model training, in the encoder awakening stage, the adapters are trained with a peak learning rate of 1e-3 using a linear warmup followed by cosine annealing decay. For the Continual fine-tuning stage, a peak learning rate of 1e-4 is used with the same schedule. {Each stage is trained for 5 epochs, and the best checkpoint is selected based on development set performance.} We also evaluate the proposed method on the speech branch of Phi-4-Multimodal~\cite{microsoft_phi4_multimodal}, a multimodal LLM from Microsoft that integrates speech and vision encoders with the Phi-4-Mini language model~\cite{abouelenin2025phi}, to verify its generalizability across different Speech-LLM backbones. All models are trained on a single NVIDIA A6000 GPU with an effective batch size of 16 (batch size $\times$ gradient accumulation steps). We use the Word Error Rate (WER) as the evaluation metric.

\subsection{Baselines}
We compare against the following baselines:
\begin{itemize}
    \item \textbf{Previous SOTA}: The best previously published WER results on each dataset using standard training data and data processing~\cite{FanSA24,ying25_wocci,shankar2026gclora}. For reference, we also report stronger results obtained using additional unlabeled data for training~\cite{wang2026gumbelbeard}.
    \item \textbf{Zero-shot}: The backbone Speech-LLM evaluated directly on the target domain without fine-tuning, serving as a measure of its zero-shot generalization capability.
    \item \textbf{Vanilla Fine-tuning}: The encoder, projector, and LLM LoRA are jointly fine-tuned on the target-domain training set using Cross-Entropy (CE) loss, as described in Section~\ref{standard_ft}. This represents the most straightforward adaptation strategy for Speech-LLMs.
    \item \textbf{Multi-stage Alignment}: The multi-stage alignment procedure from~\cite{MuSWYX25} as described in Section~\ref{multi_ft}, which sequentially fine-tunes different model components to align speech and language representations.
\end{itemize}
We follow the same data splits and processing procedures as previous SOTA works to ensure fair comparisons. To ensure a fair comparison in terms of training budget, both Vanilla and Multi-stage Alignment are trained for the same number of total epochs as the proposed method, and results are reported using the checkpoint with the best development set performance. In practice, the best checkpoint is typically obtained within the first few epochs, while additional training often leads to overfitting. {Although many fine-tuning strategies have been proposed, we select Vanilla Fine-tuning and Multi-stage Alignment~\cite{MuSWYX25} as two representative and competitive adaptation baselines from prior work.}

\begin{figure}[t]
    \centering
    \includegraphics[width=\columnwidth]{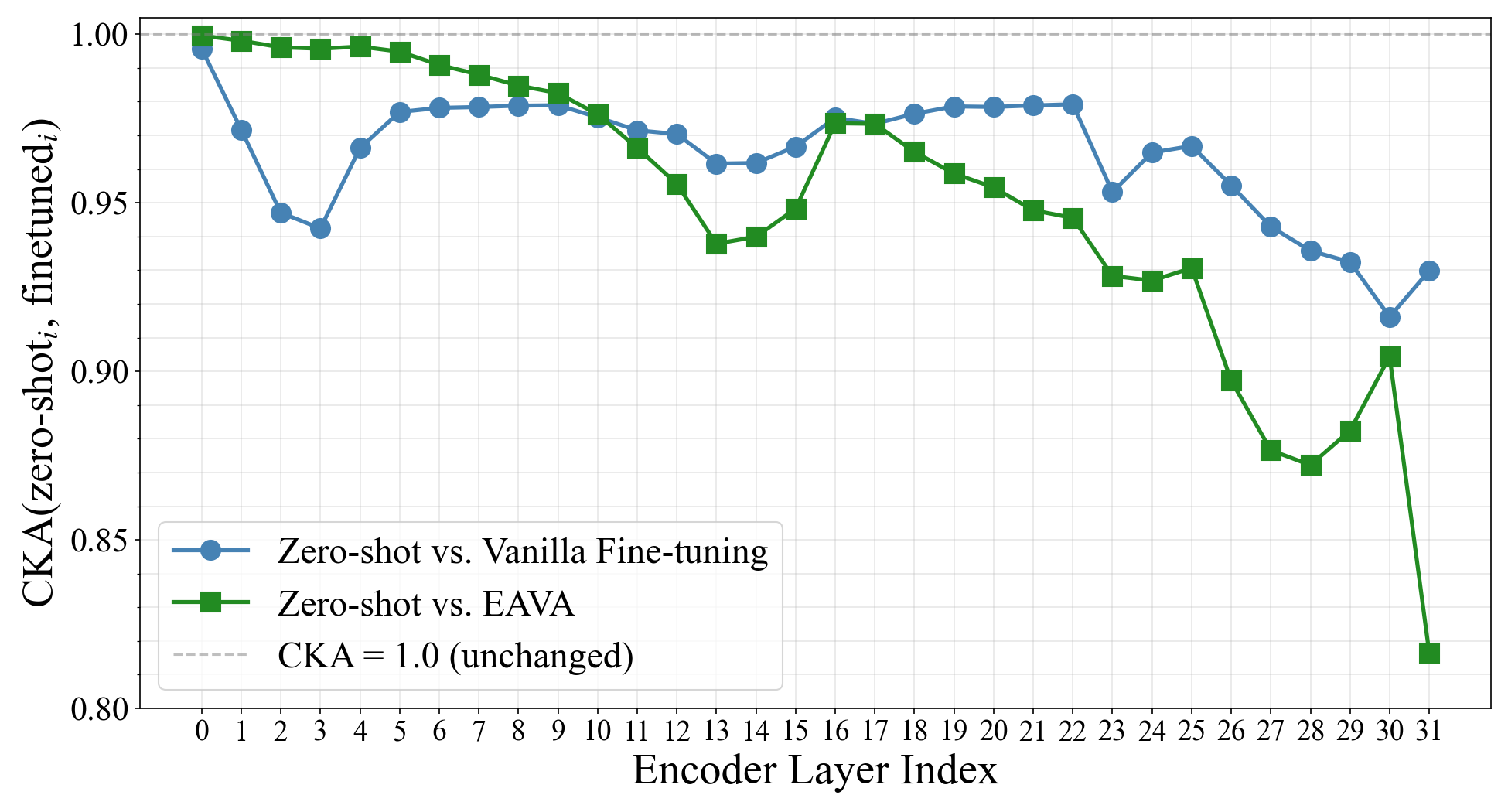}
    \vspace{-0.6cm}
    \caption{Layer-wise CKA similarity between the zero-shot and fine-tuned encoder representations on the OGI test set. Lower CKA indicates greater representational change relative to the zero-shot encoder.}
    \label{fig:cka}
    \vspace{-0.3cm}
\end{figure}

\section{Experimental Results}
\subsection{Main Results and Analysis}
Table~\ref{tab:main} presents the main results using Canary-Qwen as the backbone across all three target-domain datasets. Vanilla fine-tuning improves over zero-shot performance on all test sets, but the gains remain limited. Multi-stage alignment yields marginal improvements on OGI but degrades performance on the other two datasets, suggesting it is ill-suited for domain-adaptive fine-tuning of a well-trained Speech-LLM.

\begin{table*}[t]
\centering
\caption{Ablation study on the training procedure using Canary-Qwen as the backbone, evaluated by WER (\%) on the OGI test set. The table shows the fine-tuned modules and the number of trainable parameters at each stage. For settings with adapters, the adapter dimension is set to 64. $\dagger$ denotes the default training procedure of the proposed method.}
\vspace{-0.2cm}
\label{tab:ablation}
\begin{tabular}{lc lc c}
\toprule
Stage 1 & Params & Stage 2 & Params & WER \\
\midrule
Adapter                  & 4.3M  & --                                    & --    & 9.16  \\
Adapter$^\dagger$        & 4.3M  & Adapter+Encoder+Projector+LoRA$^\dagger$        & 842M  & \textbf{9.05}  \\
Adapter                  & 4.3M  & Encoder+Projector+LoRA                & 838M  & 9.10  \\
Adapter                  & 4.3M  & Projector+LoRA                        & 27.8M & 9.27  \\
\midrule
Adapter+Projector+LoRA   & 32M   & --                                    & --    & 9.89  \\
Adapter+Projector+LoRA   & 32M   & Adapter+Encoder+Projector+LoRA        & 842M  & 9.84  \\
Adapter+Encoder+Projector+LoRA & 842M & --                               & --    & 10.23    \\
\midrule
Full Encoder (No Adapters)             & 810M  & --                                    & --    & 13.97 \\
Full Encoder (No Adapters)          & 810M  & Encoder+Projector+LoRA (No Adapters)                & 838M  & 11.82 \\
\bottomrule
\end{tabular}
\vspace{-0.1cm}
\end{table*}

EAVA consistently outperforms all baselines across all evaluated adapter dimensions and all three test datasets. Notably, it achieves new state-of-the-art results even compared to prior works that leverage additional unlabeled training data, with the largest gains on OGI, likely because spontaneous speech from younger children represents a particularly challenging domain shift. Furthermore, larger adapters do not yield consistent improvements, possibly because the target-domain adaptation can be achieved with a relatively small number of trainable parameters, making additional adapter capacity less beneficial. Overall, the 64-dim adapter yields the most stable performance across all three datasets (test WER: OGI: 9.05\%, MyST: 8.04\%, CORAAL: 8.54\%) and is adopted as the default configuration for the remaining experiments.

To understand how each fine-tuning strategy reshapes the encoder, we compute layer-wise Centered Kernel Alignment (CKA)~\cite{kornblith2019similarity} between each fine-tuned encoder and the zero-shot encoder on the OGI test set, as it is the most challenging dataset. For each encoder layer, frame-level activations are mean-pooled over time to obtain per-utterance representations; CKA is measured at the same layer depth. Lower CKA indicates greater representational shift from the zero-shot encoder.

Figure~\ref{fig:cka} shows the layer-wise CKA diagonal. Vanilla fine-tuning produces a nearly flat curve (CKA: 0.92--0.99), indicating mild and uniformly distributed drift across all layers. Our proposed method EAVA exhibits a distinctly different profile: lower encoder layers (0--10) are almost perfectly preserved (CKA $>$ 0.97), while upper layers (23--31) diverge substantially (CKA as low as 0.82). This pattern aligns with the design rationale of our method: upper encoder layers, whose representations are passed through the projector to the LLM, benefit most from domain adaptation, while lower layers retain general acoustic features. The targeted adaptation of upper layers while preserving lower-level representations leads to more effective domain transfer.

\subsection{Ablation Study of Training Procedure}
Our proposed method consists of two stages: in Stage 1 (Encoder Awakening), lightweight adapters are inserted into each encoder layer and only the encoder adapters are trained; in Stage 2 (Continual Fine-tuning), the adapter, encoder, projector, and LLM LoRA are jointly fine-tuned. To validate this design, we conduct an ablation study on the OGI test set using Canary-Qwen as the backbone, examining the effect of varying the trainable modules in each stage (Table~\ref{tab:ablation}).

% The first four rows share the same Stage 1 configuration (adapter-only) and vary the Stage 2 components. Using Canary-Qwen backbone, even without any Stage 2 fine-tuning, the model already achieves 9.16\% WER, slightly worse than the proposed procedure. Training only a subset of modules in Stage 2 also yields slightly worse results than the full proposed configuration, suggesting that jointly fine-tuning all components in Stage 2 is beneficial.

{The first four rows share the same Stage 1 configuration (adapter-only) and vary the trainable components in Stage 2. Stage 1 performs the main encoder adaptation and already provides a strong starting point, while Stage 2 further refines the model by jointly optimizing the encoder, adapters, projector, and LLM LoRA. With the Canary-Qwen backbone, skipping Stage 2 results in a WER of 9.16\%, while the proposed procedure achieves the best performance. Fine-tuning only a subset of components in Stage 2 also leads to worse results, indicating that joint optimization provides complementary refinement beyond Stage 1.}

The next two rows examine alternative Stage 1 configurations. Training the projector and LoRA alongside the adapter in Stage 1 degrades performance compared to adapter-only training, as optimizing multiple modules simultaneously divides the training focus, making it less effective at injecting target-domain knowledge into the encoder adapters in a targeted manner. In addition, inserting adapters and training all modules jointly in a single step, without a dedicated Encoder Awakening stage, is also worse than the proposed two-stage design, demonstrating the importance of first awakening the encoder adapters before joint fine-tuning. Finally, replacing the adapters with full encoder fine-tuning in Stage 1 leads to a substantial performance drop, as it fails to preserve the encoder's pre-trained knowledge.

\begin{table}[t]
\centering
\caption{Comparison of adapter architectures in the EAVA framework using Canary-Qwen as the backbone. All adapter variants have approximately equal parameter counts (${\sim}4.2$--$4.3$M). $\dagger$ denotes the default adapter choice of our proposed method. All adapter variants show statistically significant improvements over Vanilla Fine-tuning ($p < 0.05$).}
\vspace{-0.2cm}
\label{tab:adapter}
\resizebox{\columnwidth}{!}{%
\begin{tabular}{lc cc cc cc}
\toprule
\multicolumn{1}{c}{\multirow{2}{*}{\shortstack{Adapter\\Choice}}} & \multirow{2}{*}{\shortstack{Adapter\\Params}} & \multicolumn{2}{c}{OGI} & \multicolumn{2}{c}{MyST} & \multicolumn{2}{c}{CORAAL} \\
\cmidrule(lr){3-4} \cmidrule(lr){5-6} \cmidrule(lr){7-8}
& & dev & test & dev & test & dev & test \\
\midrule
Vanilla Fine-tuning        & --   & 9.39          & 10.95          & 7.63          & 8.37          & 6.11          & 8.87 \\
\midrule
Residual Adapter~\cite{BapnaF19}$^\dagger$  & 4.3M & 8.15          & \textbf{9.05}  & 7.33          & 8.04          & 5.84          & 8.54 \\
Houlsby Adapter~\cite{houlsby2019parameter}             & 4.3M & 8.10          & 9.41           & 7.32          & \textbf{8.00} & \textbf{5.70} & \textbf{8.49} \\
LoRA~\cite{HuSWALWWC22}                        & 4.2M & \textbf{8.07} & 9.42           & \textbf{7.28} & 8.08          & 5.89          & 8.60 \\
\bottomrule
\end{tabular}%
}
\vspace{-0.2cm}
\end{table}

% \vspace{-0.3cm}
\subsection{Effect of Adapter Choice}
To investigate whether the choice of adapter architecture affects the effectiveness of EAVA, we compare three adapter variants using Canary-Qwen as the backbone, all with approximately equal parameter counts (${\sim}4.2$--$4.3$M): the Residual Adapter (dim=64) in the proposed method, Houlsby Adapter~\cite{houlsby2019parameter} (dim=32, inserted after both the attention and FFN sublayers within each encoder layer), and LoRA (rank=16, alpha=32, applied to Q, K, V, and O projections). As shown in Table~\ref{tab:adapter}, all three variants consistently outperform vanilla fine-tuning across all three datasets, demonstrating that the general design of EAVA is the primary driver of improvement, rather than the specific adapter architecture. The performance differences among the three adapter types are small, with the proposed Residual Adapter achieving the strongest result on the most challenging OGI dataset. These results confirm that EAVA is a flexible framework that is not tied to a particular adapter design.

\subsection{Cross-Dataset Transfer Analysis}
To evaluate whether the first training stage (Encoder Awakening) learns transferable domain-adaptive encoder representations, we conduct a cross-dataset transfer analysis using Canary-Qwen as the backbone. In this setting, Stage 1 is trained on one dataset, and Stage 2 is then performed separately on each target-domain dataset. As shown in Table~\ref{tab:cross_transfer}, even using mismatched data in Stage 1 outperforms vanilla fine-tuning in most cases, while using the matched target-domain data for Stage 1 generally yields the strongest results. These findings support the effectiveness of Encoder Awakening and suggest that the encoder can often still benefit from the awakening stage even when mismatched data are used.

% \begin{table*}[t]
% \centering
% \caption{Cross-dataset transfer results for Stage 1 (Encoder Awakening) using Canary-Qwen as the backbone, evaluated by WER (\%) on three target-domain datasets. In each setting, the first stage is trained on the listed dataset, and Stage 2 is performed on each target-domain dataset with 64-dim adapters.}
% \label{tab:cross_transfer}
% \begin{tabular}{lc cc cc cc}
% \toprule
% \multirow{2}{*}{Method} & \multirow{2}{*}{Dataset for Stage 1} & \multicolumn{2}{c}{OGI} & \multicolumn{2}{c}{MyST} & \multicolumn{2}{c}{CORAAL} \\
% \cmidrule(lr){3-4} \cmidrule(lr){5-6} \cmidrule(lr){7-8}
% & & dev & test & dev & test & dev & test \\
% \midrule
% Vanilla Fine-tuning & --     & 9.39 & 10.95 & 7.57 & 8.34 & 6.11 & 8.87 \\
% \midrule
% \multirow{3}{*}{EAVA (Ours)}
%                      & OGI  & \textbf{8.15} & \textbf{9.05} & 7.36 & 8.12 & 5.96 & 8.84 \\
%                      & MyST   & 8.76 & 10.06 & \textbf{7.33} & \textbf{8.04} & 6.11 & 8.98 \\
%                      & CORAAL & 8.21 & 9.45 & 7.38 & 8.10 & \textbf{5.84} & \textbf{8.54} \\
% \bottomrule
% \end{tabular}
% \end{table*}

\begin{table}[t]
\centering
\caption{Cross-dataset transfer results for Stage 1 (Encoder Awakening) using Canary-Qwen as the backbone, evaluated by test WER (\%). Stage 2 is performed on each target-domain dataset with 64-dim adapters.}
\vspace{-0.2cm}
\label{tab:cross_transfer}
\resizebox{\columnwidth}{!}{%
\begin{tabular}{lc ccc}
\toprule
Method & Training Data for Stage 1 & OGI & MyST & CORAAL \\
\midrule
Vanilla Fine-tuning & -- & 10.95 & 8.37 & 8.87 \\
\midrule
\multirow{3}{*}{EAVA (Ours)}
 & OGI    & \textbf{9.05} & 8.12 & 8.84 \\
 & MyST   & 10.06 & \textbf{8.04} & 8.98 \\
 & CORAAL & 9.45  & 8.10 & \textbf{8.54} \\
\bottomrule
\end{tabular}%
}
\end{table}

\begin{table}[t]
\centering
\caption{Comparison of loss functions used in the first stage (Encoder Awakening), evaluated by test WER (\%) on all three datasets. The backbone is Canary-Qwen, and 64-dim adapters are used. $\dagger$ denotes the default setting of our proposed method.}
\vspace{-0.2cm}
\label{tab:loss}
\begin{tabular}{lc ccc}
\toprule
Method & Loss for Stage 1 & OGI & MyST & CORAAL \\
\midrule
Vanilla Fine-tuning          & --                & 10.95 & 8.37 & 8.87 \\
\midrule
\multirow{2}{*}{EAVA (Ours)} & CTC on Encoder   & 9.60          & 8.12          & 9.05          \\
                             & CE on LLM$^\dagger$ & \textbf{9.05} & \textbf{8.04} & \textbf{8.54} \\
\bottomrule
\end{tabular}
\end{table}

\subsection{Effect of Loss Function in the Encoder Awakening Stage}
In the Encoder Awakening stage, the default supervision signal is the LLM Cross-Entropy (CE) loss for next-token prediction. We investigate whether alternative loss functions can also awaken the encoder adapters. Specifically, we attach a Connectionist Temporal Classification (CTC)~\cite{graves2006ctc} head directly to the encoder output and train the adapters using CTC ASR loss, bypassing the projector and LLM entirely during Stage 1. As shown in Table~\ref{tab:loss}, using CTC loss still outperforms vanilla fine-tuning on the OGI and MyST test sets. This observation suggests that the primary benefit comes from awakening the encoder adapters, while the choice of loss function mainly influences how this adaptation is achieved. Nevertheless, the default LLM CE loss achieves better overall performance, as it is consistent with the fine-tuning objective in Stage 2 (Continual Fine-tuning), leading to better alignment between the two stages.

\subsection{Generalizability to Other Speech-LLMs}
To verify the generalizability of the proposed method, we conduct experiments on a second Speech-LLM, Phi-4-Multimodal~\cite{microsoft_phi4_multimodal}, as shown in Table~\ref{tab:phi}. The proposed method consistently outperforms all baselines across all three datasets, demonstrating that its effectiveness is not limited to a particular backbone architecture.
% Comparing with the Canary-Qwen results, Phi-4-Multimodal achieves a lower WER under standard fine-tuning on OGI (9.92\% vs.\ 10.95\%), suggesting stronger baseline adaptation on this dataset. Accordingly, the absolute gain from EAVA over vanilla fine-tuning is smaller on OGI for Phi-4-Multimodal (0.26\% absolute) than for Canary-Qwen (1.90\% absolute), as less room for improvement remains. On MyST and CORAAL, similar trends are observed across both backbones, with improvements of 0.24\% and 0.37\% , respectively, for Phi-4-Multimodal and 0.33\% and 0.33\% for Canary-Qwen, respectively. This suggests that EAVA remains beneficial across different Speech-LLM backbones, while the magnitude of the gains depends on the characteristics of the underlying model.

Compared with the Canary-Qwen results, Phi-4-Multimodal exhibits stronger baseline adaptation on OGI under vanilla fine-tuning, resulting in a smaller margin of improvement from EAVA. Nevertheless, similar performance trends are observed across both backbones, with EAVA consistently outperforming the corresponding baselines. These results suggest that the effectiveness of EAVA extends beyond a specific Speech-LLM architecture, although the magnitude of the gains varies across backbones and datasets.

\begin{table}[t]
\centering
\caption{WER (\%) on three target-domain datasets using Phi-4-Multimodal as the backbone with 64-dim adapters. All EAVA results are statistically significant compared to baselines ($p < 0.05$).}
\vspace{-0.2cm}
\label{tab:phi}
\resizebox{\columnwidth}{!}{%
\begin{tabular}{l cc cc cc}
\toprule
\multirow{2}{*}{Method} & \multicolumn{2}{c}{OGI} & \multicolumn{2}{c}{MyST} & \multicolumn{2}{c}{CORAAL} \\
\cmidrule(lr){2-3} \cmidrule(lr){4-5} \cmidrule(lr){6-7}
& dev & test & dev & test & dev & test \\
\midrule
Zero-shot            & 18.16 & 19.23 & 9.59 &10.02 & 9.71 & 13.47 \\
Vanilla Fine-tuning & 9.29 & 9.92 & 7.46 & 8.31 & 6.24 & 9.22 \\
Multi-stage Alignment & 9.15 & 9.84 & 8.02 & 8.90 & 5.99 & 9.31 \\ 
\midrule
EAVA (Ours)     & \textbf{8.67} & \textbf{9.66} & \textbf{7.33} & \textbf{8.07} & \textbf{5.82} & \textbf{8.85} \\
\bottomrule
\end{tabular}%
}
\end{table}

%% file: conclusions.tex
\section{Conclusions}
In this work, we proposed Encoder Awakening via Adapters (EAVA), a simple yet effective method for domain-adaptive fine-tuning of Speech-LLMs. Through Encoder Awakening, target-domain knowledge is effectively injected into the encoder via lightweight adapters while preserving the encoder's pre-trained knowledge. The subsequent Continual Fine-tuning stage further adapts the entire model to the target domain. Experiments across multiple datasets and Speech-LLM backbones demonstrate that the proposed method consistently outperforms all baselines. Moreover, different adapter architectures prove effective within the EAVA framework, indicating that the general idea is the key factor rather than the specific adapter choice. We further find that larger adapters do not yield consistent improvements, likely because target-domain adaptation can be achieved with a relatively small number of trainable parameters. In addition, using mismatched data in the first stage can still awaken the encoder in most cases, indicating broader applicability of the proposed method. Furthermore, alternative supervision signals such as CTC loss can awaken the encoder adapters in the first stage in most cases, though the default LLM CE loss achieves better and more consistent overall performance. Overall, the proposed method achieves new state-of-the-art performance on all three domain-shifted datasets, surpassing even prior works that rely on additional unlabeled training data. In future work, we plan to explore more effective and efficient adaptation strategies for Speech-LLMs and other speech foundation models.